# Improvement of Text Dependent Speaker Identification System Using Neuro-Genetic Hybrid Algorithm in Office Environmental Conditions

Md. Rabiul Islam[1] and Md. Fayzur Rahman[2]

[1] **Department of Computer Science & Engineering**
**Rajshahi University of Engineering & Technology (RUET), Rajshahi-6204, Bangladesh**
*rabiul_cse@yahoo.com*

[2] **Department of Electrical & Electronic Engineering**
**Rajshahi University of Engineering & Technology (RUET), Rajshahi-6204, Bangladesh**
*mfrahman3@yahoo.com*

**Abstract**
In this paper, an improved strategy for automated text dependent speaker identification system has been proposed in noisy environment. The identification process incorporates the Neuro-Genetic hybrid algorithm with cepstral based features. To remove the background noise from the source utterance, wiener filter has been used. Different speech pre-processing techniques such as start-end point detection algorithm, pre-emphasis filtering, frame blocking and windowing have been used to process the speech utterances. RCC, MFCC, $\Delta$MFCC, $\Delta\Delta$MFCC, LPC and LPCC have been used to extract the features. After feature extraction of the speech, Neuro-Genetic hybrid algorithm has been used in the learning and identification purposes. Features are extracted by using different techniques to optimize the performance of the identification. According to the VALID speech database, the highest speaker identification rate of 100.000 % for studio environment and 82.33 % for office environmental conditions have been achieved in the close set text dependent speaker identification system.

## 1. Introduction

Biometrics are seen by many researchers as a solution to a lot of user identification and security problems now a days [1]. Speaker identification is one of the most important areas where biometric techniques can be used. There are various techniques to resolve the automatic speaker identification problem [2, 3, 4, 5, 6, 7, 8].

Most published works in the areas of speech recognition and speaker recognition focus on speech under the noiseless environments and few published works focus on speech under noisy conditions [9, 10, 11, 12]. In some research work, different talking styles were used to simulate the speech produced under real stressful talking conditions [13, 14, 15]. Learning systems in speaker identification that employ hybrid strategies can potentially offer significant advantages over single-strategy systems.

In this proposed system, Neuro-Genetic Hybrid algorithm with cepstral based features has been used to improve the performance of the text dependent speaker identification system under noisy environment. To extract the features from the speech, different types of feature extraction technique such as RCC, MFCC, $\Delta$MFCC, $\Delta\Delta$MFCC, LPC and LPCC have been used to achieve good result. Some of the tasks of this work have been simulated using Matlab based toolbox such as Signal processing Toolbox, Voicebox and HMM Toolbox.

## 2. Paradigm of the Proposed Speaker Identification System

The basic building blocks of speaker identification system are shown in the Fig.1. The first step is the acquisition of speech utterances from speakers. To remove the background noises from the original speech, wiener filter has been used. Then the start and end points detection algorithm has been used to detect the start and end points from each speech utterance. After which the unnecessary parts have been removed. Pre-emphasis filtering technique has been used as a noise reduction technique to increase the amplitude of the input signal at frequencies where signal-to-noise ratio (SNR) is low. The speech signal is segmented into overlapping frames. The purpose of the overlapping analysis is that each speech sound of the input sequence would be approximately centered at some frame. After segmentation, windowing technique has been used. Features were extracted from the segmented speech. The extracted features were then fed to the Neuro-Genetic hybrid techniques for learning and classification.

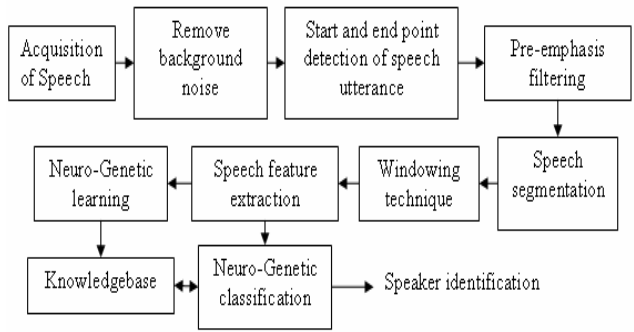

Fig. 1 Block diagram of the proposed automated speaker identification system.

## 3. Speech Signal Pre-processing for Speaker Identification

To capture the speech signal, sampling frequency of 11025 Hz, sampling resolution of 16-bits, mono recording channel and Recorded file format = *.wav have been considered. The speech preprocessing part has a vital role for the efficiency of learning. After acquisition of speech utterances, wiener filter has been used to remove the background noise from the original speech utterances [16, 17, 18]. Speech end points detection and silence part removal algorithm has been used to detect the presence of speech and to remove pulse and silences in a background noise [19, 20, 21, 22, 23]. To detect word boundary, the frame energy is computed using the sort-term log energy equation [24],

$$E_i = 10 \log \sum_{t=n_i}^{n_i+N-1} S^2(t) \qquad (1)$$

Pre-emphasis has been used to balance the spectrum of voiced sounds that have a steep roll-off in the high frequency region [25, 26, 27]. The transfer function of the FIR filter in the z-domain is [26]

$$H(Z) = 1 - \alpha \cdot z^{-1}, \quad 0 \leq \alpha \leq 1 \qquad (2)$$

Where $\alpha$ is the pre-emphasis parameter.

Frame blocking has been performed with an overlapping of 25[%] to 75[%] of the frame size. Typically a frame length of 10-30 milliseconds has been used. The purpose of the overlapping analysis is that each speech sound of the input sequence would be approximately centered at some frame [28, 29].

From different types of windowing techniques, Hamming window has been used for this system. The purpose of using windowing is to reduce the effect of the spectral artifacts that results from the framing process [30, 31, 32]. The hamming window can be defined as follows [33]:

$$w(n) = \begin{cases} 0.54 - 0.46\cos\frac{2\Pi n}{N}, & -(\frac{N-1}{2}) \leq n \leq (\frac{N-1}{2}) \\ 0, & \text{Otherwise} \end{cases} \qquad (3)$$

## 4. Speech parameterization Techniques for Speaker Identification

This stage is very important in an ASIS because the quality of the speaker modeling and pattern matching strongly depends on the quality of the feature extraction methods. For the proposed ASIS, different types of speech feature extraction methods [34, 35, 36, 37, 38, 39] such as RCC, MFCC, ΔMFCC, ΔΔMFCC, LPC, LPCC have been applied.

## 5. Training and Testing Model for Speaker Identification

Fig.2 shows the working process of neuro-genetic hybrid system [40, 41, 42]. The structure of the multilayer neural network does not matter for the GA as long as the BPNs parameters are mapped correctly to the genes of the chromosome the GA is optimizing. Basically, each gene represents the value of a certain weight in the BPN and the chromosome is a vector that contains these values such that each weight corresponds to a fixed position in the vector as shown in Fig.2.

The fitness function can be assigned from the identification error of the BPN for the set of pictures used for training. The GA searches for parameter values that minimize the fitness function, thus the identification error of the BPN is reduced and the identification rate is maximized [43].

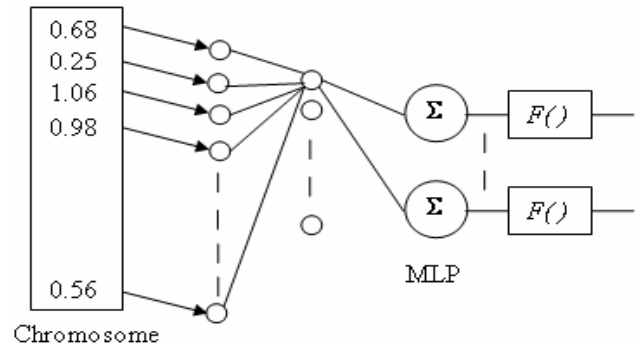

Fig.2 Learning and recognition model for the Neuro-Genetic hybrid system.

The algorithm for the Neuro-Genetic based weight determination and Fitness Function [44] is as follows:
*Algorithm for Neuro-Genetic Weight determination:*
{
$i \leftarrow 0$;
Generate the initial population $P_i$ of real coded chromosomes $C^i_j$ each representing a weight set for the BPN;

Generate fitness values $F^i_j$ for each $C^i_j \in P^i$ using the algorithm *FITGEN()*;
While the current population $P_i$ has not converged
{

Using the cross over mechanism reproduced offspring from the parent chromosome and performs mutation on offspring;

$i \leftarrow i+1$;

Call the current population $P_i$;

Calculate fitness values $F^i_j$ for each $C^i_j \in P^i$ using the algorithm *FITGEN()*;
}
Extract weight from $P_i$ to be used by the BPN;
}

*Algorithm for FITGEN():*
{Let ($\overline{I_i}$, $\overline{T_j}$), $i=1,2,\ldots\ldots N$ where $\overline{I_i} = (\overline{I_{1i}}, \overline{I_{2i}}, \ldots\ldots \overline{I_{li}})$ and $\overline{T_i} = (\overline{T_{1i}}, \overline{T_{2i}}, \ldots\ldots \overline{T_{li}})$ represent the input-output pairs of the problem to be solved by BPN with a configuration *l-m-n*.
{
Extract weights $\overline{W_i}$ from $C_i$;

Keeping $\overline{W_i}$ as a fixed weight setting, train the BPN for the *N* input instances (Pattern);
Calculate error $E_i$ for each of the input instances using the formula:

$$E_{i,} = \sum_j (T_{ji} - O_{ji})^2 \qquad (3)$$

Where $\overline{O_i}$ is the output vector calculated by BPN;
Find the root mean square $E$ of the errors $E_{i,} I = 1,2,\ldots\ldots N$
i.e. $E = \sqrt{\sum_i E_i}/N$ \qquad (4)

Now the fitness value $F_i$ for each of the individual string of the population as $F_i = E$;
}
Output $F_i$ for each $C_i, i = 1,2,\ldots\ldots P$; }

}

## 6. Optimum parameter Selection for the BPN and GA

### 6.1 Parameter Selection on the BPN

There are some critical parameters in Neuro-Genetic hybrid system (such as in BPN, gain term, speed factor, number of hidden layer nodes and in GA, crossover rate and the number of generation) that affect the performance of the proposed system. A trade off is made to explore the optimal values of the above parameters and experiments are performed using those parameters. The optimal values of the above parameters are chosen carefully and finally find out the identification rate.

#### 6.1.1 Experiment on the Gain Term, $\eta$

In BPN, during the training session when the gain term was set as: $\eta_1 = \eta_2 = 0.4$, spread factor was set as $k_1 = k_2 = 0.20$ and tolerable error rate was fixed to 0.001[%] then the highest identification rate of 91[%] has been achieved which is shown in Fig.3.

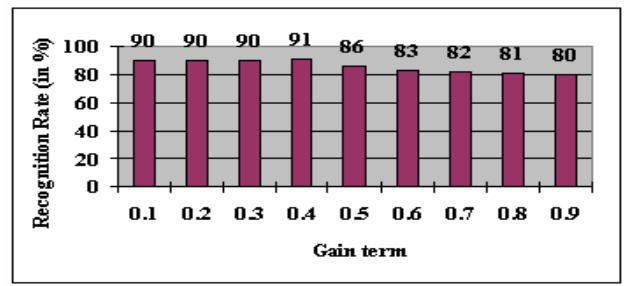

Fig. 3 Performance measurement according to gain term.

#### 6.1.2 Experiment on the Speed Factor, *k*

The performance of the BPN system has been measured according to the speed factor, *k*. We set $\eta_1 = \eta_2 = 0.4$ and tolerable error rate was fixed to 0.001[%]. We have studied the value of the parameter ranging from 0.1 to 0.5. We have found that the highest recognition rate was 93[%] at $k_1 = k_2 = 0.15$ which is shown in Fig.4.

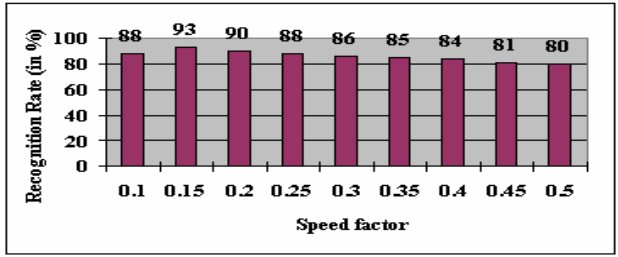

Fig. 4 Performance measurement according to various speed factor.

### 6.1.3 Experiment on the Number of Nodes in Hidden Layer, $N_H$

In the learning phase of BPN, We have chosen the hidden layer nodes in the range from 5 to 40. We set $\eta_1 = \eta_2 = 0.4$, $k_1 = k_2 = 0.15$ and tolerable error rate was fixed to 0.001[%]. The highest recognition rate of 94[%] has been achieved at $N_H = 30$ which is shown in Fig.5.

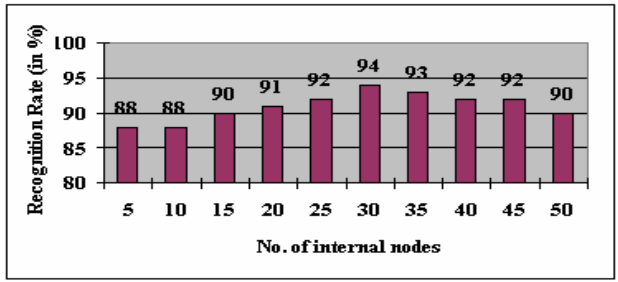

Fig. 5 Results after setting up the number of internal nodes in BPN.

### 6.2 Parameter Selection on the GA

To measure the optimum value, different parameters of the genetic algorithm were also changed to find the best matching parameters. The results of the experiments are shown below.

### 6.2.1 Experiment on the Crossover Rate

In this experiment, crossover rate has been changed in various ways such as 1, 2, 5, 7, 8, 10. The highest speaker identification rate of 93[%] was found at crossover point 5 which is shown in the Fig.6.

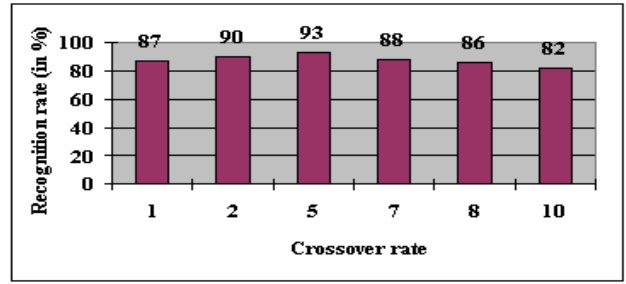

Fig. 6 Performance measurement according to the crossover rate.

### 6.2.2 Experiment on the Crossover Rate

Different values of the number of generations have been tested for achieving the optimum number of generations. The test results are shown in the Fig.7. The maximum identification rate of 95[%] has been found at the number of generations 15.

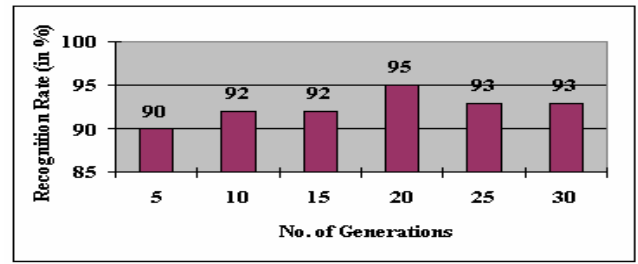

**Fig.7** Accuracy measurement according to the no. of generations.

## 7. Performance Measurement of the Text-Dependent Speaker Identification System

VALID speech database [45] has been used to measure the performance of the proposed hybrid system. In learning phase, studio recording speech utterances ware used to make reference models and in testing phase, speech utterances recorded in four different office conditions were used to measure the accurate performance of the proposed Neuro-Genetic hybrid system. Performance of the proposed system were measured according to various cepstral based features such as LPC, LPCC, RCC, MFCC, ΔMFCC and ΔΔMFCC which are shown in the following table.

**Table 1:** Speaker identification rate (%) for VALID speech corpus

| Type of environments | MFCC | Δ MFCC | ΔΔ MFCC | RCC | LPCC |
|---|---|---|---|---|---|
| Clean speech utterances | 100.00 | 100.00 | 98.23 | 90.43 | 100.00 |
| Office environments | 80.17 | 82.33 | 68.89 | 70.33 | 76.00 |

| | speech utterances | | | | | |
|---|---|---|---|---|---|---|

Table 1 shows the overall average speaker identification rate for VALID speech corpus. From the table it is easy to compare the performance among MFCC, ΔMFCC, ΔΔMFCC, RCC and LPCC methods for Neuro-Genetic hybrid algorithm based text-dependent speaker identification system. It has been shown that in clean speech environment the performance is 100.00 [%] for MFCC, ΔMFCC and LPCC and the highest identification rate (i.e. 82.33 [%]) has been achieved at ΔMFCC for four different office environments.

## 8. Conclusion and Observations

The experimental results show the versatility of the Neuro-Genetic hybrid algorithm based text-dependent speaker identification system. The critical parameters such as gain term, speed factor, number of hidden layer nodes, crossover rate and the number of generations have a great impact on the recognition performance of the proposed system. The optimum values of the above parameters have been selected effectively to find out the best performance. The highest recognition rate of BPN and GA have been achieved to be 94[%] and 95[%] respectively. According to VALID speech database, 100[%] identification rate in clean environment and 82.33 [%] in office environment conditions have been achieved in Neuro-Genetic hybrid system. Therefore, this proposed system can be used in various security and access control purposes. Finally the performance of this proposed system can be populated according to the largest speech recognition database.

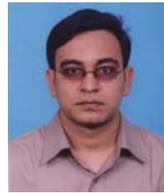

**Md. Rabiul Islam** was born in Rajshahi, Bangladesh, on December 26, 1981. He received his B.Sc. degree in Computer Science & Engineering and M.Sc. degrees in Electrical & Electronic Engineering in 2004, 2008, respectively from the Rajshahi University of Engineering & Technology, Bangladesh. From 2005 to 2008, he was a Lecturer in the Department of Computer Science & Engineering at Rajshahi University of Engineering & Technology. Since 2008, he has been an Assistant Professor in the Computer Science & Engineering Department, University of Rajshahi University of Engineering & Technology, Bangladesh. His research interests include bio-informatics, human-computer interaction, speaker identification and authentication under the neutral and noisy environments.

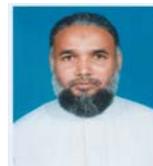

Md. Fayzur Rahman was born in 1960 in Thakurgaon, Bangladesh. He received the B. Sc. Engineering degree in Electrical & Electronic Engineering from Rajshahi Engineering College, Bangladesh in 1984 and M. Tech degree in Industrial Electronics from S. J. College of Engineering, Mysore, India in 1992. He received the Ph. D. degree in energy and environment electromagnetic from Yeungnam University, South Korea, in 2000. Following his graduation he joined again in his previous job in BIT Rajshahi. He is a Professor in Electrical & Electronic Engineering in Rajshahi University of Engineering & Technology (RUET). He is currently engaged in education in the area of Electronics & Machine Control and Digital signal processing. He is a member of the Institution of Engineer's (IEB), Bangladesh, Korean Institute of Illuminating and Installation Engineers (KIIEE), and Korean Institute of Electrical Engineers (KIEE), Korea.

**IJCSI**